\documentclass{PoS}

\newcommand{\reseteqnum}{\setcounter{equation}{0}}

\newcommand{\ovl}[1]{\overline{#1}}
\newcommand{\wt}[1]{\widetilde{#1}}

\newcommand{\eqn}[1]{(\ref{#1})}
\newcommand{\p}{\partial}
\newcommand{\bpsi}{{\overline{\psi}}}

\newcommand{\pslash}{p\kern-1ex /}
\newcommand{\Dslash}{{\cal D}\kern-1.5ex /}
\newcommand{\tr}{{\rm tr}}

\newcommand{\msbar}{\overline{\rm MS}}

\title{Non-perturbative renormalization of $N_f=2+1$ QCD with Schr\"odinger
functional scheme}

\ShortTitle{Non-perturbative renormalization of $N_f=2+1$ QCD with
SF scheme}

\author{\speaker{Yusuke Taniguchi}  for PACS-CS collaboration%
\\
Institute of Physics, University of Tsukuba,
Tsukuba, Ibaraki, 305-8571, Japan \\
	E-mail: \email{tanigchi@het.ph.tsukuba.ac.jp}}

\abstract{
We present a preliminary result of $N_f=2+1$ QCD running coupling in
Schr\"odinger functional scheme.
We adopted Iwasaki gauge action and non-perturbatively improved Wilson
fermion action with clover term.
We use seven renormalization scales to cover from low energy
to high energy perturbative region and three lattice spacings to take
the continuum limit at each scale.

A scaling behabior of the step scaling function is discussed together
with its renoralization group flow in the continuum.
We argue on introduction of the physical scale through the Sommer scale
$r_0$.
}

\FullConference{The XXVI International Symposium on Lattice Field Theory \\
		 July 14 - 19, 2008\\
		 Williamsburg, Virginia, USA}

\begin{document}

\section{Introduction}

Strong coupling and quark masses constitute fundamental parameters of
QCD part in the standard model.
One of the important task of lattice QCD is to determine these
parameters from inputs at low energy scale.
Hadron masses, meson decay constants and quark potential quantities
are adopted as physical inputs and QCD running coupling and quark masses
should be given.
These results may be compared with independent evaluation from high
energy inputs, which  may uncover existence of systematic error if there
is.

In a course of evaluating these fundamental parameters in the Lagrangian
we need a process of renormalization in some scheme.
It is now recognized that systematic deviation due to perturbative
renormalization is large for quark mass and also for running coupling
at low energy region.
Non-perturbative renormalization is essential for this work.
Among several non-perturbative schemes on the lattice the Schr\"odinger
functional (SF) scheme
\cite{Luscher:1992an,Luscher:1993gh,Sint:2000vc,DellaMorte:2004bc}
has an advantage that systematic errors can be unambiguously controlled:
A unique renormalization scale is introduced through the box size to
reduce the lattice artifact and a large range of the renormalization
scale can be covered by the step scaling function (SSF) technique.
The latter virtue matches our purpose to make comparison with high
energy inputs.

For the SF scheme we need to start with evaluation of the running
coupling in order to introduce the renormalization scale.
It is strongly expected that three light quarks should involve in
running of the coupling at low energy scale, where non-perturbative
effects becomes large.
So introduction of $u, d, s$ quarks would be important for
non-perturbative running of the coupling.
This report presents preliminary results of our calculation for the
running coupling in $N_f=2+1$ QCD with SF scheme.
The physical scale shall be introduced through the Sommer scale
$r_0$ evaluated independently by the CP-PACS collaboration with light
$N_f=2+1$ configuration \cite{Aoki:2008sm}.

\reseteqnum
\section{Schr\"odinger functional formalism and action}

The Schr\"odinger functional is given as a field theory in a finite box
of size $L^4$ with a Dirichlet boundary condition at temporal boundary.
For QCD the Dirichlet boundary condition is set for spatial component of
the gauge link
\begin{eqnarray}
U_k(x)|_{x_0=0}=\exp\left(a C_k\right),\quad
U_k(x)|_{x_0=T}=\exp\left(a C_k'\right),\quad
C_k^{(')}=
\frac{i}{L}\pmatrix{\phi_1^{(')}\cr&\phi_2^{(')}\cr&&\phi_3^{(')}\cr}
\label{eqn:boundary}
\end{eqnarray}
and quark fields
\begin{eqnarray}
\psi(x)|_{x_0=0}=\psi(x)|_{x_0=T}=0,\quad
\bpsi(x)|_{x_0=0}=\bpsi(x)|_{x_0=T}=0.
\end{eqnarray}
Under a lenient condition it is proved that the tree level gauge
effective action has a global minimum around a background field $V_\mu$
which is uniquely given by the boundary fields \eqn{eqn:boundary}.
On the other hand the fermionic mode is shown to have a mass gap, with
which we are able to define mass independent scheme in the chiral limit
without any extrapolation.
The renormalization scale is given only by the box size $L$.

We adopt the renormalization group improved gauge action of Iwasaki type
\begin{eqnarray}
S_g=
 \frac{\beta}{N}\sum_{{ C}\in{ S}_0}W_0({ C},g_0^2)
{\rm Re\ }\tr\left(1-P({ C})\right)
+\frac{\beta}{N}\sum_{{ C}\in{ S}_1}W_1({ C},g_0^2)
{\rm Re\ }\tr\left(1-R({ C})\right),
\end{eqnarray}
where $S_0$ and $S_1$ are sets of oriented plaquettes and rectangles.
The weight factor $W_{0/1}$ is given to cancel the $O(a)$
contribution from the boundary
according to \cite{Takeda:2003he,Takeda:2004xh}.
The boundary improvement coefficients are set to tree level value
$c^P_{\rm{t}}=1$ and $c^R_{\rm{t}}=3/2$, which is shown to give better
scaling behavior than one loop value \cite{Takeda:2004xh,Murano}.
We take the same values for boundary link \eqn{eqn:boundary} as in the
previous work of the Alpha collaboration
\cite{Luscher:1993gh,DellaMorte:2004bc}.

We used the improved Wilson fermion action with clover term
\begin{eqnarray}
&&
S_f 
=a^4\sum_{x}\bpsi\left(D_W+m_0\right)\psi,
\quad
D_W=\frac{1}{2}\left(\gamma_\mu\left(\nabla_\mu+\nabla_\mu^*\right)
-a\nabla_\mu^*\nabla_\mu\right)
-c_{\rm SW}\frac{1}{4}\sigma_{\mu\nu}P_{\mu\nu}.
\end{eqnarray}
The improvement coefficient $c_{\rm SW}$ is given non-perturbatively in
a polynomial form \cite{Aoki:2005et}
which covers $1.9\le\beta\le12.0$.
We notice that there is a contribution from the boundary to cancel the
$O(a)$ effect there
\begin{eqnarray}
S_{{ O}(a)}&=&
a^3\sum_{\vec{x}}\left(\wt{c}_t-1\right)
\left(\bpsi(\vec{x},1)\psi(\vec{x},1)+\bpsi(\vec{x},T-1)\psi(\vec{x},T-1)
\right),
\end{eqnarray}
for which the one loop value \cite{Aoki:1998qd} is taken
$\wt{c}_{t}=1-0.00881(28)g_0^2$.
We set twisted periodic boundary condition in spatial direction
$\psi(x+L\hat{k})=e^{i\theta}\psi(x)$ with $\theta={\pi}/{5}$
\cite{Luscher:1993gh,DellaMorte:2004bc}.

The renormalized gauge coupling in the SF scheme is defined as a
coefficient of the effective action $\Gamma[V_\mu]$ at the global
minimum.
For numerical simulation we take derivative in terms of a parameter
$\eta$ introduced in the background field $\phi_i$ and define the SF
coupling as \cite{Luscher:1993gh}
\begin{eqnarray}
\frac{1}{\ovl{g}^2(L)}=
\frac{1}{k}\left.\frac{\p\Gamma[V_\mu]}{\p\eta}\right|_{\eta=0},
\end{eqnarray}
where $k$ is a normalization coefficient evaluated at tree level.

\reseteqnum
\section{Our strategy}

The goal of our project for the running coupling is to derive the
renormalization group invariant (RGI) scale $\Lambda_{\rm QCD}$ in a
unit of the Sommer scale $r_0$.
The RGI scale $\Lambda$ is scheme dependent and is defined as follows
for the SF scheme
\begin{eqnarray}
\Lambda_{\rm SF}=
\frac{1}{L}\left(b_0\ovl{g}(L)\right)^{-\frac{b_1}{2b_0^2}}
\exp\left(-\frac{1}{2b_0\ovl{g}(L)}\right)
\exp\left(-\int_{0}^{\ovl{g}(L)}dg
\left(\frac{1}{\beta(g)}+\frac{1}{b_0g^3}-\frac{b_1}{b_0^2g}\right)\right),
\label{eqn:lambda}
\end{eqnarray}
where $\ovl{g}(L)$ is a renormalized coupling in SF scheme at a
scale $L$ and $\beta(g)$ is renormalization group function
($\beta$-function) with its perturbative expansion coefficients
\begin{eqnarray}
\beta(g)=-g^3\left(b_0+b_1g^2+b_2g^4+\cdots\right).
\end{eqnarray}
Derivation of the RGI scale is given by the following steps in the SF
scheme.

(i) We start by calculating the SSF $\Sigma$ on the lattice at several
box sizes and lattice spacings.
The SSF gives a relation between the renormalized couplings when the
renormalization scale is changed by factor two
$\Sigma(u,a/L)=\left.\ovl{g}^2(2L)\right|_{u=\ovl{g}^2(L)}$
\cite{Luscher:1993gh,DellaMorte:2004bc},
where the scale is given by 
the renormalized coupling $\ovl{g}^2(L)$ and discretization error by
$a/L$.
Taking the continuum limit $\sigma(u)=\lim_{a/L\to0}\Sigma(u,a/L)$ and
performing a polynomial fit we have a full non-perturbative running of
the coupling in a discretized manner.

(ii) In the second step we define a reference scale $L_{\rm max}$
through a fixed value of renormalized coupling $\ovl{g}^2(L_{\rm max})$.
The value of $\ovl{g}^2(L_{\rm max})$ is rather ambiguous if it is well
in low energy region.
We then start from $L_{\rm max}$ and follow non-perturbative RG flow
through the SSF into high energy region.
After $n\sim8$ iterations the scale $L=2^{-n}L_{\rm max}$ is already in
perturbative region where discrepancy between perturbative and
non-perturbative RG running is negligible.

(iii) Substituting $\ovl{g}^2(L)$ and $L=2^{-n}L_{\rm max}$ given in the
above into \eqn{eqn:lambda} and evaluating the integral
with three loops $\beta$-function in the SF scheme \cite{Bode:1999sm} we
get the RGI scale $\Lambda_{\rm SF}L_{\rm max}$ in terms of the
reference scale.

(iv) In the last step we need the physical input $r_0$ measured in an
independent large scale simulation at some lattice spacing $a$.
The reference scale should also be measured at the same lattice spacing
to give a ratio $r_0/L_{\rm max}$.
The requirement for the lattice spacing and the reference scale is that
magnitude of the lattice artifact $a/r_0$ and $a/L_{\rm max}$ should be
kept small.
Multiplying these factors we get the RGI scale $\Lambda_{\rm SF}r_0$ in
terms of the Sommer scale.
Transformation into the $\msbar$ scheme is given exactly at one loop
$\Lambda_{\ovl{\rm MS}}=2.612\Lambda_{\rm SF}$.


\reseteqnum
\section{Step scaling function}

We adopted seven renormalized couplings to cover from the weak coupling
region $\ovl{g}^2=1.001$ to strong region $\ovl{g}^2=3.418$ separated
approximately by twice the renormalization scale.
For each coupling we used three boxes $L/a=4, 6, 8$ to take the
continuum limit.

HMC algorithm is adopted for two flavours and RHMC algorithm for the third
flavour, all of which are taken to be the common mass.
We adopted CPS++ code and modified for SF formalism.
For machines we make use of T2K, PACS-CS and PC cluster Kaede at
University of Tsukuba, T2k and SR11000 at University of Tokyo and PC
cluster RSCC at Riken.

We start by tuning the value of $\beta$ and $\kappa$ to reproduce the
same renormalized coupling at each box sizes keeping the PCAC mass to
zero, where the PCAC relation is defined in terms of the improved axial
current with non-perturbative improvement coefficient
\cite{Kaneko:2007wh}.
We notice that distribution of inverse of the coupling $1/\ovl{g}^2$
turned out to be smooth Gaussian even at the lowest energy scale
\cite{Murano}.
This is contrary to the standard Wilson gauge action
\cite{DellaMorte:2004bc} and we need no re-weighting.

The renormalized coupling $\ovl{g}^2(2L)$ at larger scale is modified
perturbatively in order to cancel
deviation of the PCAC mass from zero and that of the renormalized
coupling $\ovl{g}^2(L)$ from a fixed value
\cite{DellaMorte:2004bc,Bode:1999sm,Sint:1995ch}.
A part of $O(a)$ error is canceled at one loop level with coefficients
given in Ref.~\cite{Takeda:2003he}.
In the end we get the $O(a)$ improved SSF on the lattice.

\begin{figure}
 \begin{center}
  \includegraphics[width=5.6cm]{sigma.continuum.eps}
  \includegraphics[width=6cm]{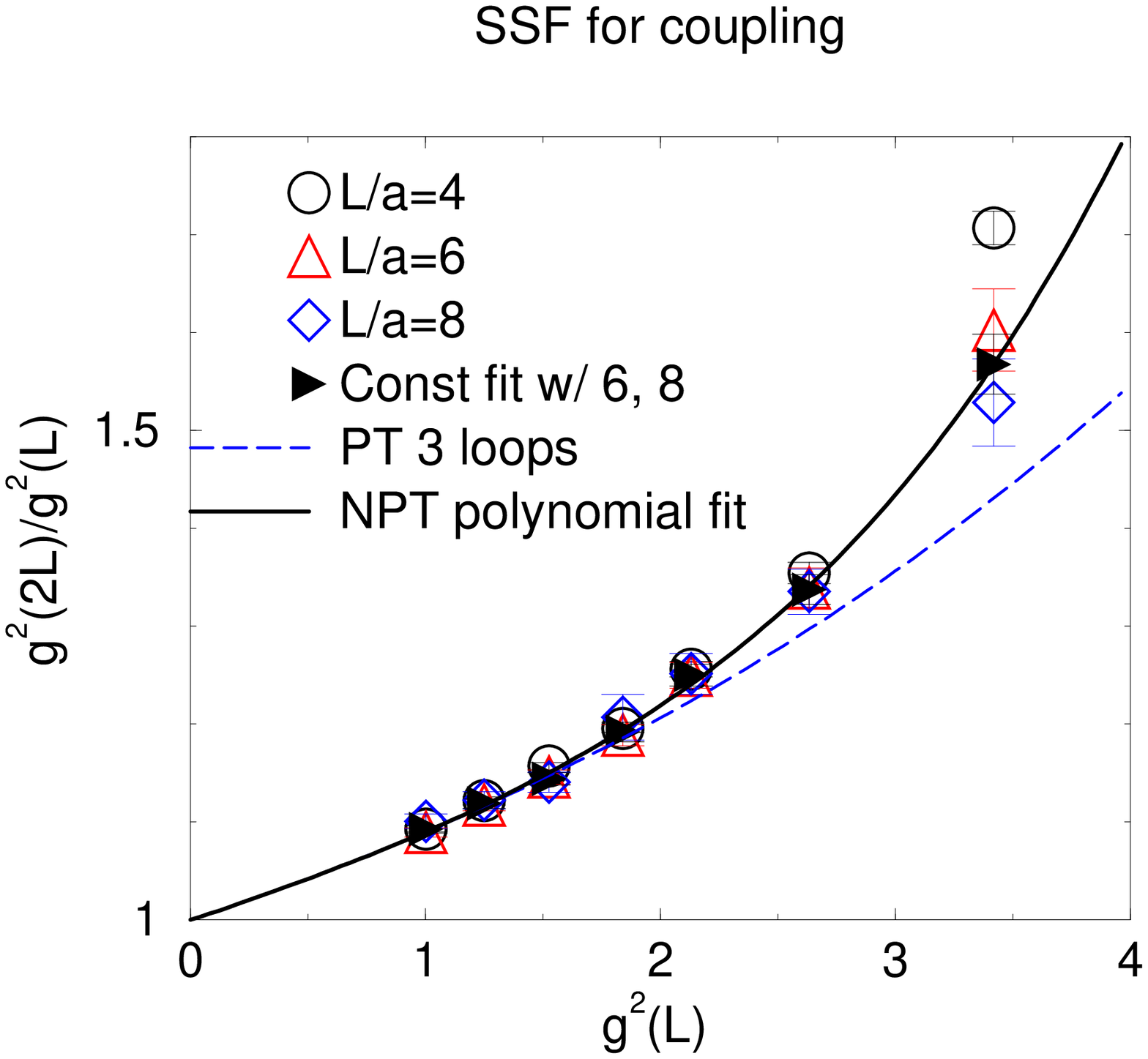}
  \caption{SSF on the lattice with its continuum extrapolation at each
  renormalization scale (left).
  RG flow of the SSF 
  (right).
  }
  \label{fig:SSF}
 \end{center}
\end{figure}

\begin{figure}
 \begin{center}
  \includegraphics[width=6cm]{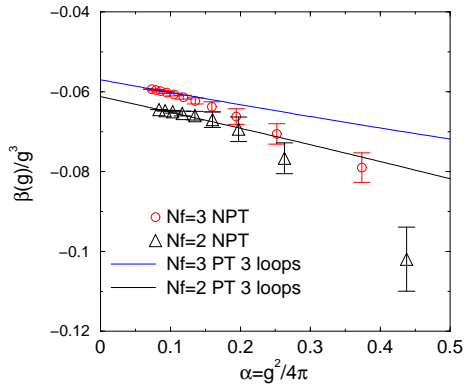}
  \caption{Non-perturbative $\beta$-function for $N_f=3$ and $2$ QCD.
  }
  \label{fig:npt-beta}
 \end{center}
\end{figure}

Preliminary result is plotted in figure \ref{fig:SSF}.
The left panel shows scaling behavior of the SSF at each renormalization
scale, which turned out to be good except at the strongest coupling.
We performed three types of continuum extrapolation:
constant extrapolation with finest two (filled symbols) and three data
points (open symbols) and linear extrapolation with three data (open
circles).
As is plotted in the figure they are consistent with each other except
at the strongest.
We adopted the continuum fit with finest two lattice spacings as our
preliminary continuum value.
In the right panel the RG running of the SSF is plotted.
We divide the SSF with the coupling $\ovl{g}^2(L)$ to get better
resolution.
Polynomial fit of the continuum SSF to sixth order
\begin{eqnarray}
\sigma(u)=u+s_0u^2+s_1u^3+s_2u^4+s_3u^5+s_4u^6
\label{eqn:poly}
\end{eqnarray}
is plotted (solid line) together with the three loop perturbative
running (dotted line).
We used one and two loop values for $s_0$ and $s_1$ in the fit.
From the polynomial form of the SSF we derive the non-perturbative
$\beta$-function of the $N_f=3$ QCD, which is plotted in figure
\ref{fig:npt-beta}.
The $\beta$-function of $N_f=2$ QCD is reproduced from data of the Alpha
collaboration \cite{DellaMorte:2004bc} for comparison.

\reseteqnum
\section{Introduction of physical scale}

From a independent simulation of PACS-CS collaboration at $\beta=1.90$
\cite{Aoki:2008sm} preliminary value of the Sommer scale is given as
$a/r_0=0.131(35)$ in the chiral limit.
Evaluation of the strong coupling in the SF scheme at the same $\beta$
in $4^4$ box gives $\ovl{g}^2(L_{\rm max})=4.695(23)$ in the chiral limit.
We adopt this coupling as a definition of $L_{\rm max}$.
\footnote{Unfortunately simulation in $6^4$ box gives
$\ovl{g}^2(L)=6.71(16)$, which exceeds our largest coupling
$\ovl{g}^2=5.35(10)$ for the SSF.}

Starting from $u_{\rm max}=4.695(23)$ we iterate non-perturbative RG
flow eight times according to polynomial fit \eqn{eqn:poly} and
substitute the result into \eqn{eqn:lambda}.
In the end we get $\Lambda_{\rm SF}L_{\rm max}=0.238(19)$.
Multiplying $a/r_0$ and $L_{\rm max}/a=4$ we change the reference scale
to $r_0$ and we have $\Lambda_{\rm SF}r_0=0.45(12)$.
The large error mainly comes form systematic error of $a/r_0$ during
chiral extrapolation of strange quark mass.
We need few more points to take the massless limit in a rigid way.
In order to give $\Lambda_{\ovl{\rm MS}}$ in a unit of MeV we also need
to check validity of $r_0=0.5$ fm in the chiral limit.

\reseteqnum
\section{Step scaling function for quark mass}

Since we are calculating the PCAC mass with the $O(a)$ improved axial
current 
it is possible to derive renormalization
factor for the pseudo scalar density as a byproduct, from which we can
extract non-perturbative running of the quark mass.
However the scaling behavior of the pseudo scalar density was shown
to be bad even perturbatively \cite{Kurth:2002rz} under the
inhomogeneous boundary gauge field \eqn{eqn:boundary} and twist factor
$\theta=\pi/5$ in spatial direction.

\begin{figure}
 \begin{center}
  \includegraphics[width=5.3cm]{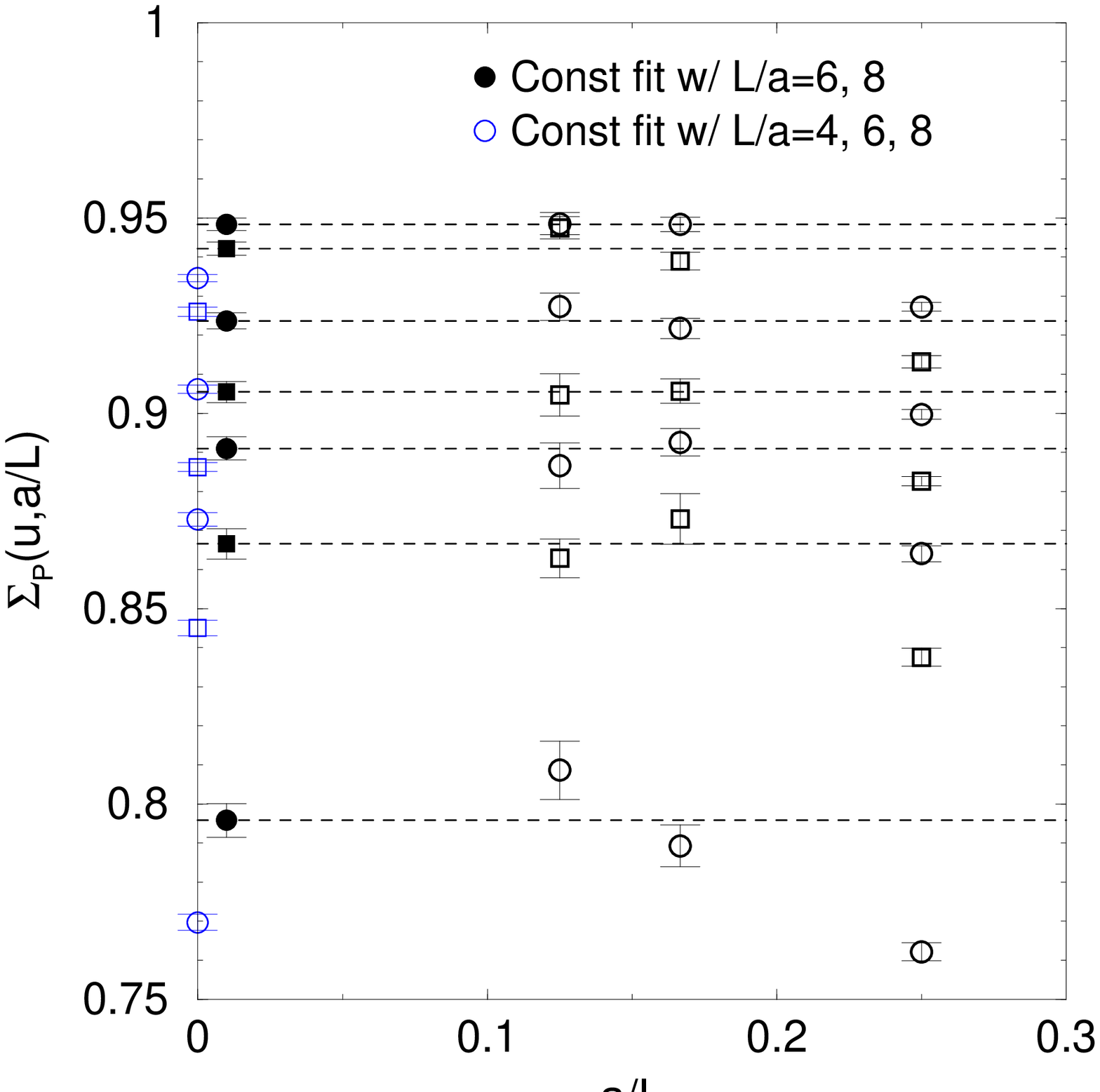}
  \includegraphics[width=5.3cm]{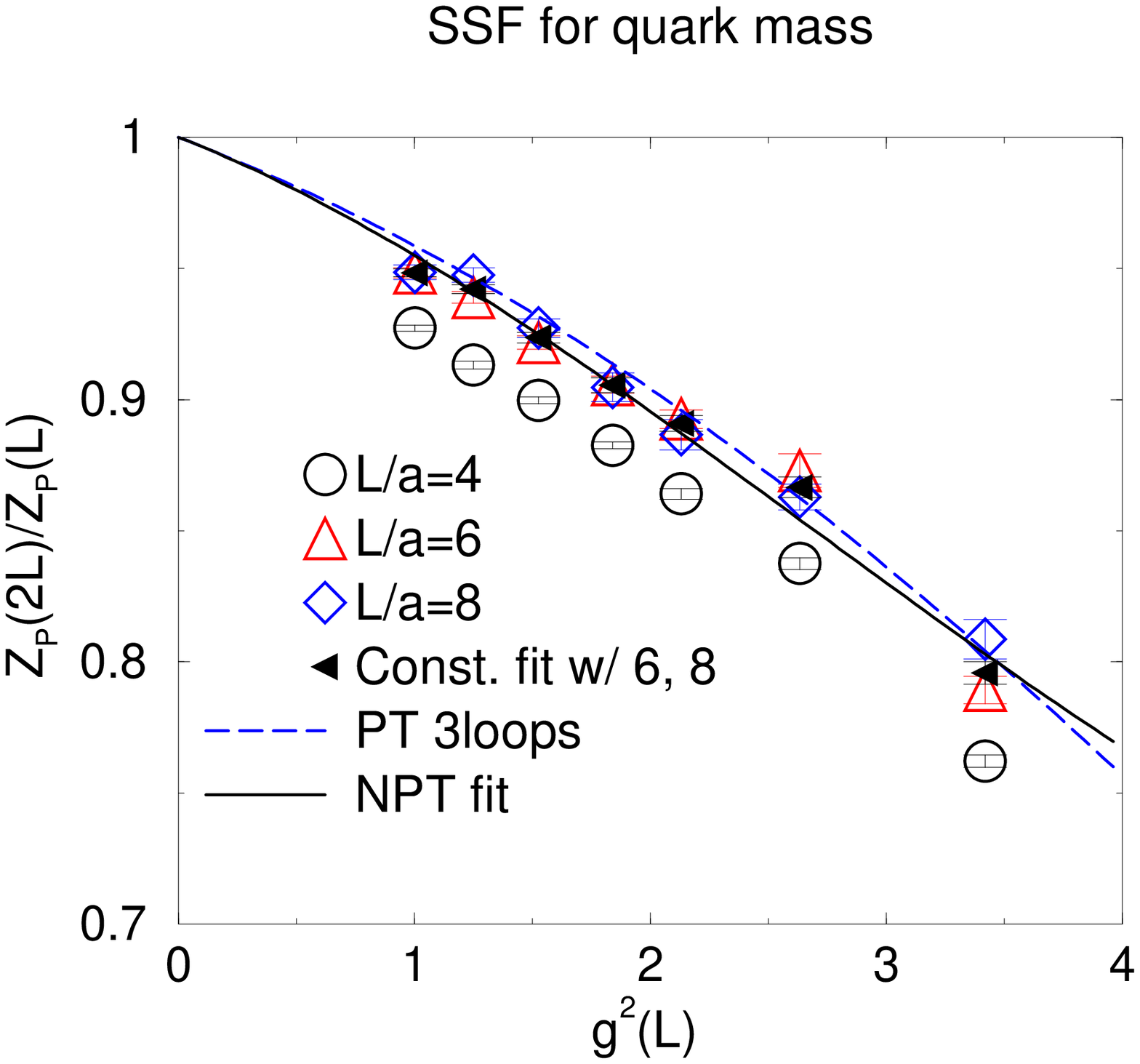}
  \caption{SSF of the pseudo scalar density on the lattice with its
  continuum extrapolation at each renormalization scale (left).
  RG flow of the SSF (right).
  }
  \label{fig:SSFzp}
 \end{center}
\end{figure}

In this report we define the pseudo scalar density renormalization
factor as
\begin{eqnarray}
Z_P(g_0,L/a)=
\frac{f_P(x_0=L/2)_{({\rm lattice})}^{({\rm tree})}}
{\sqrt{3\left(f_1\right)_{({\rm lattice})}^{({\rm tree})}}}
\frac{\sqrt{3f_1}}{f_P(x_0=L/2)},
\end{eqnarray}
where propagators $f_P$ and $f_1$ are given in \cite{Sint:2000vc}.
We expect cancellation of $O(a)$ effect at tree level dividing by tree
level propagator on the lattice.
The SSF on the lattice is given by
\begin{eqnarray}
\Sigma_P\left(u,\frac{a}{L}\right)
=\left.\frac{Z_P(g_0,2L/a)}{Z_P(g_0,L/a)}\right|_{\ovl{g}^2(L)=u,m=0}.
\end{eqnarray}
The result is plotted in figure \ref{fig:SSFzp}, which shows a rather
bad scaling behavior.
Although we take the continuum limit with finest two lattice spacings in
this report, we may need finer lattice spacings to reduce systematic
uncertainty, which seems to be unrealistic for computational cost.

\reseteqnum
\section{Conclusion}

We present a preliminary result for the $N_f=2+1$ QCD running coupling
in the mass independent SF scheme in the chiral limit.
We used seven scales to cover from low energy to high energy
region and three lattice spacings to take the continuum limit at
each scale.

Tuning of $\beta$ and $\kappa$ has been completed to fix seven scales
in the massless limit.
We are now evaluating the SSF at the finest lattice spacings.
Our preliminary result shows a good scaling behavior except at the
lowest energy scale, for which we may need finer lattice spacing to take
the continuum limit.
In order to evaluate $\Lambda_{\ovl{\rm MS}}$ precisely we need to
derive $a/r_0$ in the chiral limit in more rigid way.
Main source of the systematic error is an extrapolation of the strange
quark mass and we are planning to perform simulation at different
parameters for $r_0$.
We also need to check validity of $r_0=0.5$ fm in the chiral limit
before we evaluate $\Lambda_{\ovl{\rm MS}}$ in terms of MeV.

The scaling of the pseudo scalar density SSF is
rather bad under inhomogeneous background gauge field.
We may need better setup with vanishing background field,
which we are planning as a next step.

This work is supported in part by Grants-in-Aid of the Ministry
of Education, Culture, Sports, Science and Technology-Japan
 (NOs. 18740130, 18104005).


\begin{thebibliography}{99}
\bibitem{Luscher:1992an}
M. L\"uscher, R. Narayanan, P. Weisz and U. Wolff,
Nucl.\ Phys.\ B {\bf 384} (1992) 168.

\bibitem{Luscher:1993gh}
M. L\"uscher, R. Sommer, P. Weisz and U. Wolff,
Nucl.\ Phys.\ B {\bf 413} (1994) 481.

\bibitem{Sint:2000vc}
S.~Sint,
Nucl.\ Phys.\ Proc.\ Suppl.\  {\bf 94} (2001) 79
and references there in.

\bibitem{DellaMorte:2004bc}
M.~Della Morte, R.~Frezzotti, J.~Heitger, J.~Rolf, R.~Sommer and U.~Wolff
 [ALPHA Collaboration],
  Nucl.\ Phys.\ B {\bf 713} (2005) 378.

\bibitem{Aoki:2008sm}
  S.~Aoki {\it et al.}  [PACS-CS Collaboration],
  arXiv:0807.1661 [hep-lat].

\bibitem{Takeda:2003he}
  S.~Takeda, S.~Aoki and K.~Ide,
  Phys.\ Rev.\  D {\bf 68} (2003) 014505.

\bibitem{Takeda:2004xh}
  S.~Takeda {\it et al.},
  Phys.\ Rev.\ D {\bf 70} (2004) 074510.

\bibitem{Murano}
K.~Murano, S.~Aoki, S.~Takeda and Y.~Taniguchi,
in this proceedings.

\bibitem{Aoki:2005et}
  S.~Aoki {\it et al.}  [CP-PACS Collaboration],
  Phys.\ Rev.\  D {\bf 73} (2006) 034501.

\bibitem{Aoki:1998qd}
  S.~Aoki, R.~Frezzotti and P.~Weisz,
  Nucl.\ Phys.\  B {\bf 540} (1999) 501.

\bibitem{Bode:1999sm}
  A.~Bode, P.~Weisz and U.~Wolff  [ALPHA collaboration],
  Nucl.\ Phys.\  B {\bf 576} (2000) 517
  [Erratum-ibid.\  B {\bf 600} (2001\ ERRAT,B608,481.2001) 453].

\bibitem{Kaneko:2007wh}
  T.~Kaneko, S.~Aoki, M.~Della Morte, S.~Hashimoto, R.~Hoffmann and R.~Sommer,
  JHEP {\bf 0704} (2007) 092.

\bibitem{Sint:1995ch}
S.~Sint and R.~Sommer,
Nucl.\ Phys.\ B {\bf 465} (1996) 71.

\bibitem{Kurth:2002rz}
  S.~Kurth,
  arXiv:hep-lat/0211011.

\end{thebibliography}
\end{document}